\documentstyle[twocolumn,prc,aps,epsf,epsfig]{revtex}
 \begin{document}
 
\newcommand{\be}{\begin{eqnarray}}
\newcommand{\ee}{\end{eqnarray}}
\twocolumn[\hsize\textwidth\columnwidth\hsize\csname@twocolumnfalse\endcsname
\title{ Jet Fragmentation due to a Quark/Diquark Pick-up \\
 in High Energy Heavy Ion Collisions
}

\author{ J.Casalderrey-Solana and  E.V. Shuryak  }
\address{
Department of Physics and Astronomy,\\
State University of New York, 
Stony Brook NY 11794-3800, USA
}

\date{\today}
\maketitle

\begin{abstract}
We propose a model aimed at explaining jet quenching, large azimuthal
asymmetry
and baryon/meson ratio at large transverse momenta  $p_t=2-10\, GeV$
observed at RHIC. Its main point is that a QCD string can be cut
by
 matter quarks and/or diquarks before its natural breaking.
We model the early quark/diquark production via QCD sphalerons.
\end{abstract}
\vspace{0.1in}
]
\newpage

1.{\bf Observations.} 
The field of heavy ion collisions entered the new era with first
experiments at Relativistic Heavy Ion Collider (RHIC),  which
revealed a number of new phenomena. While
the totality of data for most secondaries, with $p_t< 2 \,\, GeV/c$, agree
with a picture of strong collective explosion of matter close to
equilibrium,
and is well described  by relativistic hydrodynamics \cite{hydro},
 for $p_t> 2 \,\, GeV/c$ the regime changes. Based on pp data
 it has been anticipated that
this region can be described perturbatively, by a
standard
parton model, with  modest modifications   due to initial and
final
state interaction. (Such ideology was implemented e.g.
in popular  event generator HIJING.)  Instead, RHIC experiments
at $p_t=2-10\, GeV$ have found that:
(i) hadron yields,  relative  to pp $or$
parton model, are smaller by the `` jet quenching
factor'' $Q\sim 1/3$; (ii) the  azimuthal asymmetry is unexpectedly
 large, with 
 $v_2=<cos(2\phi)>\sim .1$   for mid-peripheral collisions; (iii)
the  baryon/meson ratio is unexpectedly large, $\sim 1$, well above that
in the usual 
jet fragmentation; (iv) a clear link between the two last points is
seen from the fact that $v_2$ for baryons ($p,\Lambda$)
is larger than for mesons ($\pi,K$).
In this paper we propose a mechanism which may account for all
these observations.

2.{\bf Recent literature.} 
A simple theoretical argument made by one of us \cite{Shu_v2} is
 that in any model of jet quenching by $absorption$  is limited
by the
 regime of surface emission. Its consequence is the
{\em geometrical limit}: $v_2(b)< v^*_2(b)$ where the r.h.s.
is uniquely related to  the shape   of
 the almond-shaped nuclear
overlap region at impact parameter $b$. However,  
 as emphasized in \cite{Shu_v2},  STAR data seem to
 {\em exceed} the geometric limit \cite{comment1}, which rules out
 all
purely absorptive models.  

It was suggested by
Lin and Ko \cite{Lin:2002rw} and Voloshin and Molnar \cite{Voloshin}
that 
quark coalescence into
hadrons  enhances $v_2$, and $v_{baryon}/v_{meson}\approx 3/2$.
The coalescence 
has been further discussed in Refs\cite{Fries:2003vb,Greco:2003mm}.
It is concluded that the coalescence of 2 or 3 $hard$ partons is very
improbable. The coalescence of {\em soft thermalized partons} 
produces 
thermal-like distribution of hadrons in the rest frame of each matter
cell, reproducing basically the Cooper-Frie formula
used in all hydro papers. How many partons participates in 
coalescence seem to be nearly irrelevant.
A coalescence of
{\em  hard-soft} kind  may  enhance the
 baryon/meson ratio. 
However, due to jet quenching the hard partons only contribute to
hadronic spectra if they are produced at the surface and move outward. 
The soft ones, to be lifted by flow, have to wait a significant
  time of   $\sim 10\, fm/c$, at which point hard
ones are too far away. Our model to be introduced below
includes a hard-soft coalescence at very early time, $<1.5\, fm/c$.
 Hard and soft partons are not at the same place, but are connected by
a string.

3.{\bf The model.} Unlike
models mentioned above, we focus on the matter-induced
modification of the jet fragmentation process. Our first new point is that 
although the outgoing hard quark (or gluon) leaves the system promptly, 
the QCD string (or 2 strings) is still extended behind
and is crossing the excited matter. The string helps to explain the
timing problem  mentioned above, among other ones.
The second new point is that the
QCD sphalerons can provide  promptly quarks or diquarks,
conveniently concentrated on thin and expanding spheres \cite{OCS}.    

  New mechanism of string breaking
  must act {\em before} (in the jet center of mass frame) than both
 (i) the usual spontaneous string-breaking, as described by the Lund-model
\cite{BookA}; and (ii) the
 perturbative gluon radiation from a parton. (
 Due to
 strongly falling $p_t$ spectra,  even modest
energy loss makes the contribution irrelevant.)
 An approximate expression for the time of both phenomena (i) and (ii)
combined \cite{B.Kopeliovich} can be written as a condition
\be \tau < \tau_{string \, breaking}\approx {p_t \over \sigma +p_t^2/45} \ee
where $\sigma=(.44 \, \, GeV/c)^2$ is the usual string tension. 

The ``string cutting'' is the main idea of the model, while the
sphalerons used as a description of matter at early time is 
 admittedly an extreme one, and can possibly be later
 combined/replaced by another one.
It was 
chosen due to its simple geometrical rules, and also
to maximize  both 
(i) the radiative jet quenching and (ii) quark and diquark pick-up rate.
We hope the reader will find it reasonable to go to the extreme
 in the first exploratory
study,  since so many other
models we and others tried
have failed.

Theory of the {\em QCD sphalerons} is discussed in detail in
 \cite{OCS}. In brief, they are
  unstable classical soliton-like objects with masses $\sim 3
 \, \, GeV/c$, excited from the part of the vacuum wave function
under the topological barrier by  high energy collisions. 
 
For discussion of their production
in hadronic collisions in experiments see e.g. \cite{SZ_03} and
 references therein. 

  Like jets, sphalerons are produced
in (semi)-hard $gg$ collisions and thus have the same distribution of the
 origination points in the transverse plane.
 Once
produced, they evolve into expanding spheres, also moving with a speed
of light.
 The
 radiative energy loss in a single crossing of a parton
through the sphaleron was calculated in \cite{SZ_quenching}.
Using its results we estimated that this generate quenching
of a parton by about an order of magnitude,
provided the time is shorter than $\tau< \tau_{sp}$.
After that time we stop the model.

The 
rules of the model thus are as follows:
(i) If a jet goes through the sphaleron  sphere, it is 
eliminated; (ii) If a jet escapes all the sphalerons but its string
 is crossed by a sphere of one of them, the string is cut by $q$
 pickup.
 (iii) With
 a probability $P_{qq}$ a diquark instead of a quark is picked up.
The  parameters we will use below are (i) the sphaleron density
$dN_{sp}/dy=200$ in central AuAu;  (ii) $P_{qq}=1$; (iii) 
the lifetime of the process $\tau_{sp}=1.5 \, fm/c$;

Further evolution of a $\bar q-string-q$ ($q-string-qq$)
systems is done following the Lund model.
As they are produced 
  earlier compared with the creation of pairs in the usual
  fragmentation, those
have relatively small invariant masses.

4.{\bf Fragmentation}
in the Lund model (e.g. in  PYTHIA \cite{PYTHIA,SJ}
event generator) describe a
 gluon jet  as a pair of 
 strings that stretches from it to the forward-backward-moving
 quarks \cite{BookA}.  The usual treatment 
 only fulfills the Lund Area Law on average 
which is inadequate for small masses  
dominated by the
few-body decays.
Fragmentation  of the low energy quark jets has been studied 
only recently \cite{AH} in this
framework, passing the tests provided by the BES collaboration \cite{BES}. 

Unfortunately, the corresponding fragmentation
of the low mass gluon strings have not yet been studied. 
Since the production of relevant particles  at RHIC is dominated by
gluon jets, we have to address the issue.
Additionally, we found that
 the issue of the shape of the gluon string is very important. It in turn
is related to string-string attraction,
 first introduced by Montvay \cite{Montvay}. If the tension of the
 double string is less than twice that of a single one,
 $\sigma_g/(2\sigma_q)\equiv r<2$, the minimal energy configuration of
 a $qg\bar{q}$ jet system have shapes shown in Fig.1. 
 The value of $r$ is not yet fixed from data. 
 T. Sj\"{o}strand \cite{Sjp} concluded only
 that the this ratio should be $r> 1.5$ in order to describe the 3-jet
 data,
 we use a value  $r =1.8 $ \cite{comm2}.

The
 junction  moves according to the forces produced by strings
 along the direction of the gluon, with the velocity  $v=r/2\approx 0.9$,
and the whole fragmentation is simplified in its rest frame.
 In this frame, if no breaking occurs, the partons will oscillate in
 the direction of the strings in a  yo-yo motion.
 The turning time of the gluon (in this frame) 
is $t_{c}=Eg_j/\sigma_g$ where $Eg_j$ is the energy in the junction rest frame.

 The string configuration just described
 can interact with quarks located at 
 sphaleron spheres.
 If a string can interact with several sphalerons,
 we  assume that the  fragmentation is determined by the one that cuts
 the q-strings closest to the junction, in its rest frame.
(As  the gluonic piece of a string is very short
and extended outwards, the cutting happens predominantly in the quark
 part
as shown in Fig.1 by the horizontal dashed lines.)

Once the pick up   happens, one is
 left with a system with low invariant mass $M$. For example,
hadrons of $p_t= 3 -4\, \, GeV/c$ come from jet subsystems with $M\sim 3
 \, GeV$,
 which is precisely where few-body fragmentation starts to be important.
One may work out the complete
exclusive fragmentation distribution for the low energy gluons,
which we have not yet done. 

 At this energy, quark jet fragmentation is dominated by three and
 four body decays \cite{AH}. However, if in the interaction with the sphaleron
 the string picks up a diquark, a two body decay
or a single string breaking shown in Fig.1 would be enough.
So,
 whenever a baryon is produced, the fragmentation  is described by a
 combination
of a two body decay plus the standard fragmentation  described by
 standard fragmentation functions. 
In the former case, the energy-momentum conservation together with the
 linear potential of the string determine uniquely the four momentum
 of the two particles produced. 
The kinematics alone ensure
 that the effect disappears with the
 increasing  $p_t$ of the gluon.

In our simulation, with $r=1.8$, the 2-body fragmentations
 die out around $p_t\sim 8\, Gev/c$, which corresponds to the
 length of the gluonic string of 1 $fm$ in the junction rest frame: longer
  strings should  break by the usual mechanism \cite{comm3}. 
The typical time and position of pick up is such that in the junction 
rest frame, the turning time of the gluon defined above is 
smaller than the pick up time. 
This means that the string works as a slingshot,
allowing the gluon to give all its energy to the baryon or pion.
 By requiring that the invariant masses of the two subsystems after the spontaneous breaking correspond to a nucleon and a $\pi$, we obtain two possible four momenta (corresponding to the two possible cuts). Surprisingly the previous remark makes the hadron that absorbs the gluon string to move in opposite direction to the initial one because in the moment of the pick up the gluon is moving back-wards. This characteristic is maintained when we translate to the original frame. The outcome is that we obtain a very energetic particle (the one that does not absorb the gluon string) with fraction of the three momentum of the gluon close to one. In fact, as the second hadron moves in the opposite direction, this fraction can be even slightly bigger than one. 

\begin{figure}
\begin{center}
\epsfig{figure=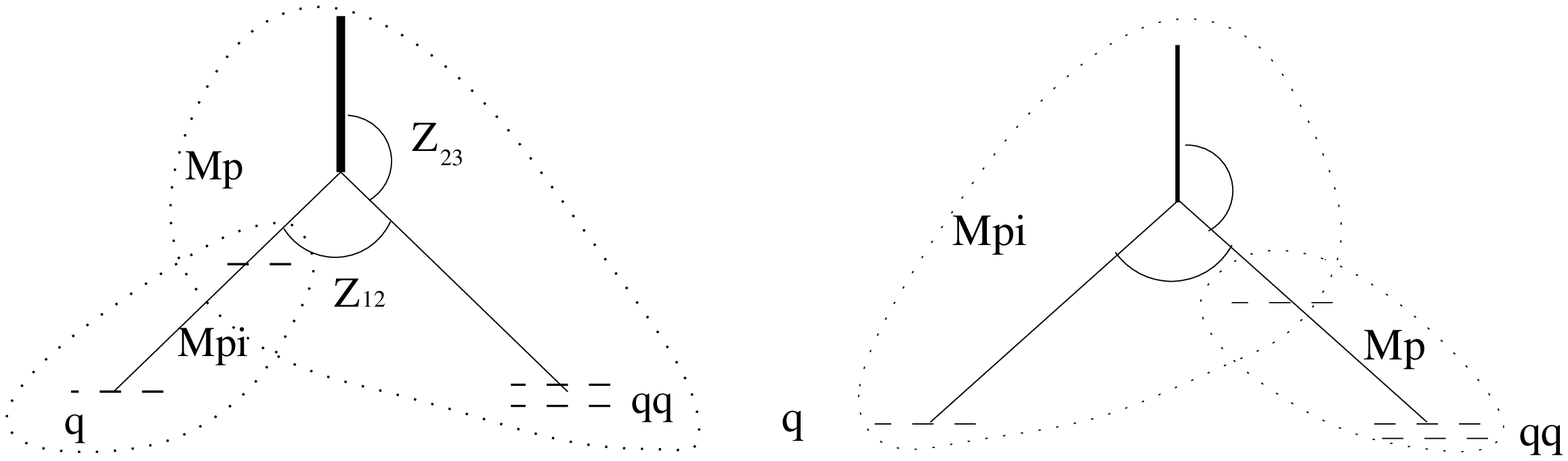,width=80mm}
\caption{String configuration of a $qg\bar{q}$ event. The pick up from the sphaleron produces a q and a diquark. The two permitted spontaneous breakings are shown}
\end{center}
\end{figure}

The two possible ways of breaking correspond to boosting either the meson or the nucleon produced. The requirement of producing physical masses is more easily fulfilled when the $\pi$ does not absorb the gluon string (as it is expected, given its small mass). So, in many cases only one fragmentation is possible. When both are possible, we have assumed that the two cuts happen with equal probability. In principle this does not have to be true and phase space and dynamical (area law suppression) considerations should be taken into account. We have also assumed that only $\pi$\'{}s or nucleons are produced. A more realistic model should also consider other channels

Summarizing this point:
 a gluon fragmentation function at moderate $p_t$ contains a part
in which only two hadrons are generated. Those are harder than the
 usual ones
 because one of the particles takes almost all the momentum of 
the initial gluon.
 Due to string dynamics, it can even happen that the three momentum of
 one of the particles is even bigger than the momentum of the gluon.

5.{\bf
The results} follow from numerical simulation of non-central AuAu
collisions at mid-rapidity, in which we produce quark and gluon jets
according to standard nuclear shapes and the structure
functions of the parton model, reproducing pp data.
All trajectories are traced, some jets are quenched by sphalerons; from those that escape some have their strings cut off, with remnants fragmenting as described above.

  The {\em pion quenching factor} is a combination of absorption due to
  quenching and enhancement due to modified fragmentation. The results
shown in Fig 2 are in agreement with the data from PHENIX 
\cite{dEnterria} for $p_t=3-7\, GeV/c$. 
 The pick up mechanism disappears
for $p_t > 8 \,\, GeV/c$ where we are only left with the strong
absorption produced by the sphalerons.

{\em The baryon/meson ratio} close to 1
 is not trivial to obtain:
 even though the only process we consider always generates one nucleon
 and one $\pi$, the boosts that these particles obtain are different. 
In general it is easier to generate high energy $\pi$\'{}s than
 protons.
 These $\pi$\'{}s come from the string piece that does not include 
the gluon string; so it will be easier to generate a particle with
 small mass from this cut. 
The other cut is suppressed by kinematic reasons. 
So, if we fix the $p_t$ of the gluons we will always 
obtain more $\pi$ than protons. 
However, we observe that when the nucleon is boosted, 
it carries a bigger fraction of the momentum. Finally,
 when we convolute yields
 with the cross section we obtain the ratio 
shown in Fig.3. We found
 that the value decreases with $p_t$ as it is expected, although  slowly.

  We have set the probability of picking up diquarks $P_{qq}=1$.
 As the mechanism of production of particles links the nucleon and
 $\pi$ production, the ratio of those particles coming from this
 mechanism will be independent of $P_{qq}$. $P_{qq}$ determines,
 however, the strength of the effect. If we reduce it, the pick up of
 quarks (that we have not included in the version reported here, 
for simplicity)
 would start to play a role.

\begin{figure}
\begin{center}
\epsfig{figure=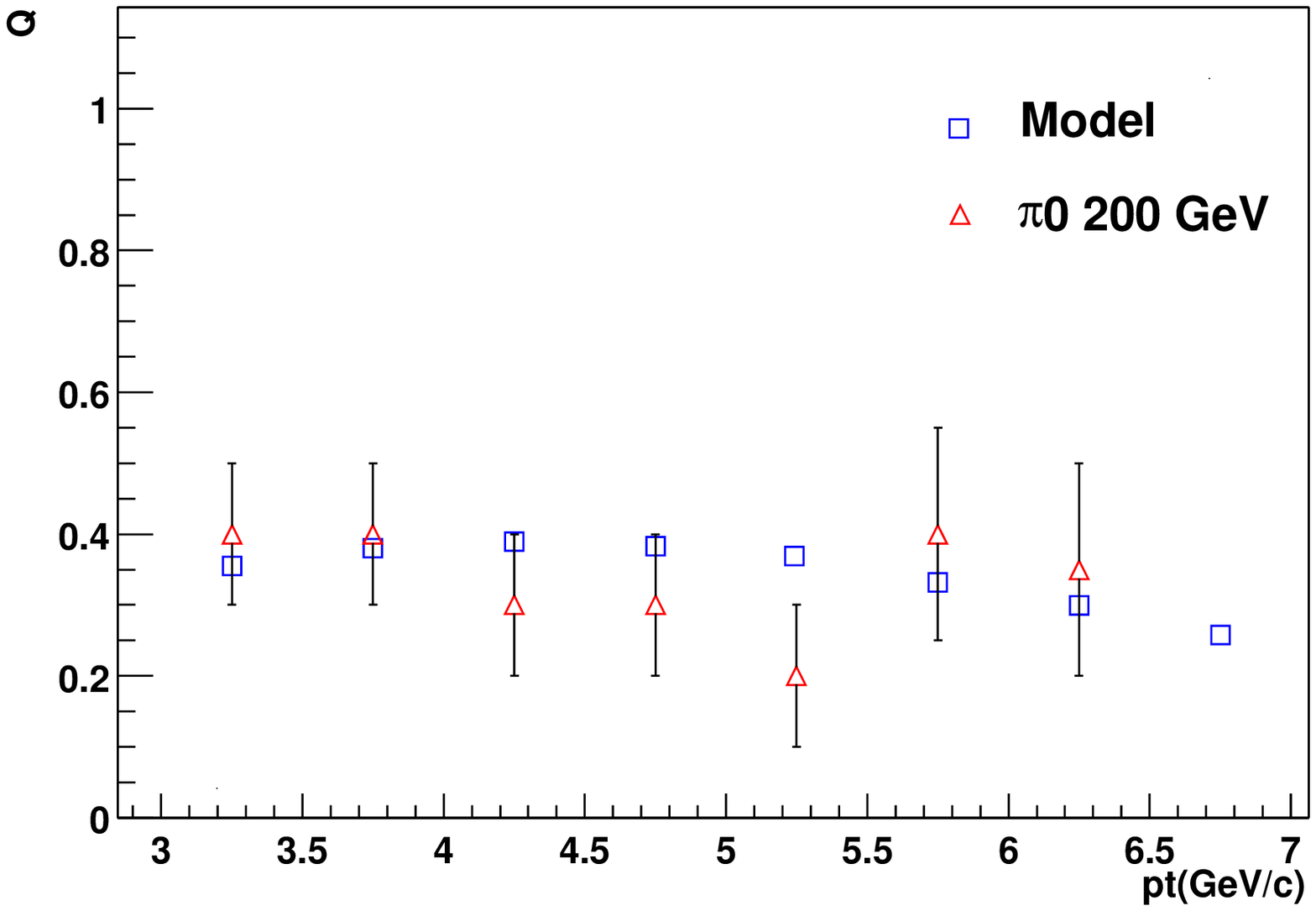,width=80mm}
\epsfig{figure=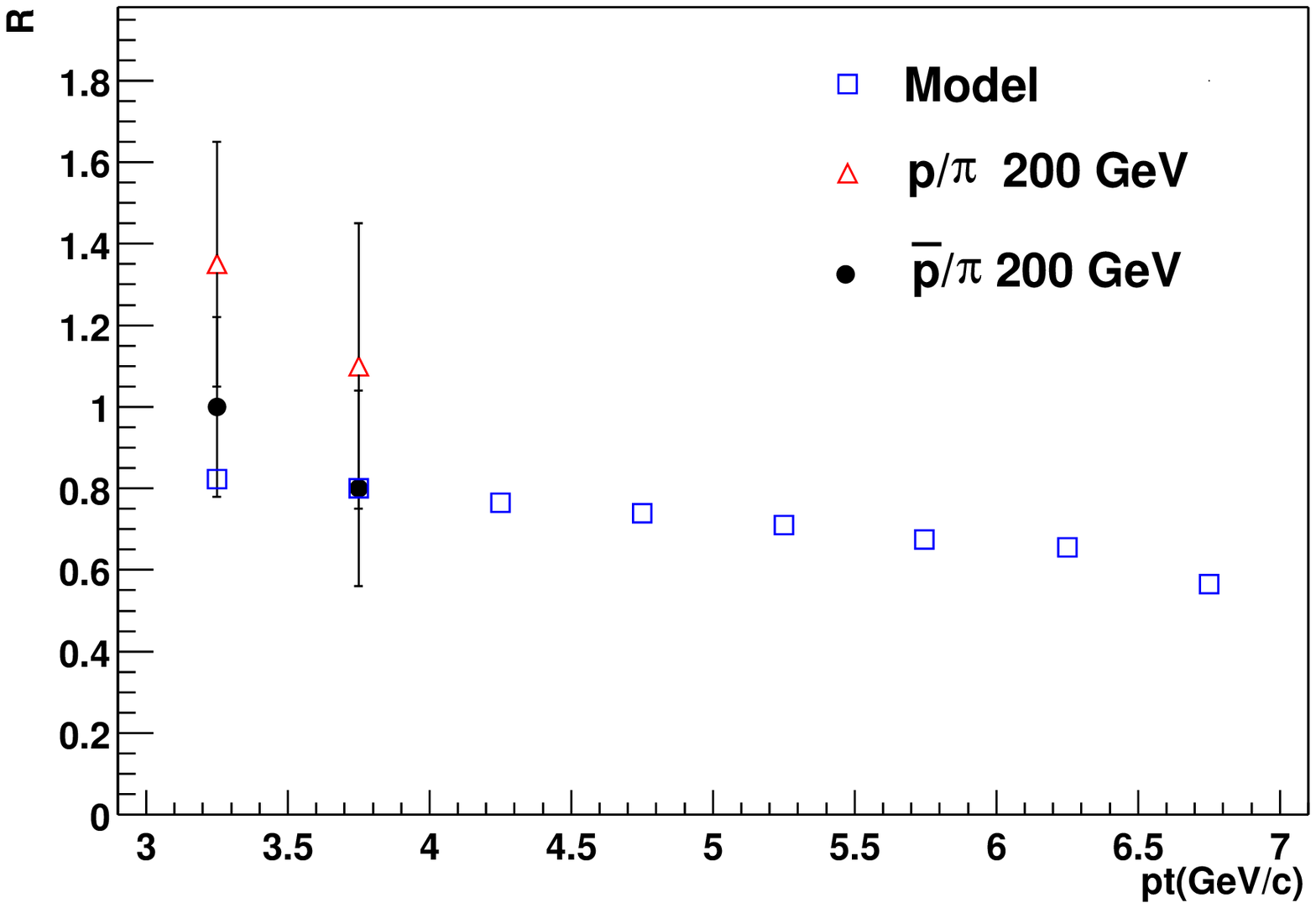,width=80mm}
\caption{(a) Quenching factor for pions and (b) $p / \pi$ ratio
versus the transverse momentum at
 impact parameter $b=6fm$. Experimental values
are PHENIX results in the 10-20 \% cetrality bin, $\sqrt
{s}=200\, GeV/c$,
   from (a)
 \protect\cite{dEnterria} and (b) from  \protect\cite{JJtalk}. }
\end{center}
\end{figure}

 As already explained, the production of
 only two $\pi$ would not be dominant because the invariant mass is
 too big. Other channels could be important, for example the
 production of $\rho$ instead of $\pi$.   The introduction of $\rho$ does not lead directly to the reduction of the $p/\pi$ ratio. We have assumed in the simulation that all strings produce a $\pi$, but this is not necessarily true. So by introducing any other channel we will reduce the number of $\pi$ even though all of them will generate high energy nucleons.

{\em Azimuthal asymmetry} follows from underlying geometry in a
non-trivial way.
 The absorption reduces the 
contribution of jets produced in the center of the almond,
 pushing the production of particles that escape 
toward the surface halo of the nuclei (see Fig.3(b)).This
is precisely the problem of all absorptive models. 

However, the density of sphalerons is smaller in the halo
and  the  (di)quark
  pick up (which boosts the fragmentation) is more
effective for more central jets.
 The outcome of both processes is that the emission is dominated by a
relatively thin layer, see Fig.3(a). As it is shaped
  approximately as the surface of the overlap region of two hard
  spheres, the model produces
 values of $v_2$ close to the geometric limit.
  
\begin{figure}
\begin{center}
\epsfig{figure=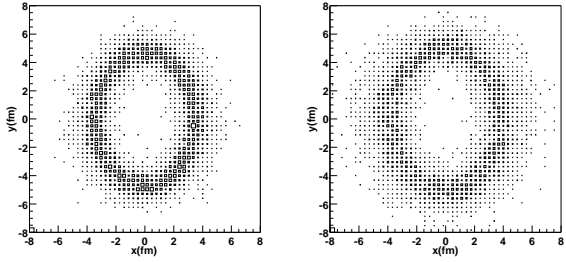,width=80mm}
\caption{Transverse plane distribution of the jets that escape the nuclei ($b=6 \, fm$ along the x axis) from pure absorption (right) and after the pick up (left). We can observe the smaller contribution of the halo.}
\end{center}
\end{figure}

Fig 3(b) shows the distribution of jets in the transverse plane (at $b=6 \,{fm}$) that escape the absorption. The big contribution from the halo reduces the value of $v_2$ to $0.049 \pm 0.001$ . Fig 3(a) shows the same distribution for $\pi$\'{}s when the pick up is considered. The contribution of the halo is reduced, yielding a
 $v_2$ at $b=6 \, fm$ 
of $0.076 \pm 0.004$ and $0.080 \pm 0.004$
for pions and nucleons. These values are found to be approximately independent of $p_t$ 
The experimental values at the same impact parameter reported by STAR  \cite{AT} are  $0.12 \pm 0.02$ 
and  $0.13 \pm 0.01$,(all charged) for   $\sqrt s=130$ and $200A\, GeV/c$.

Although after the pick up our results are lower than the experimental values, they are still  larger than
the results of  pure absorption (in spite of having a quenching factor ten times bigger).
 What is important, 
 with the pick-up  mechanism we have been able to reduce  
the contribution of the halo to the final production.
 Note also,
that the nucleon asymmetry is somewhat larger than the pion one, although
not much.

6.{\bf Conclusions.} We have presented a model in which the medium interacts not only with  the partons produced in the collision, but also with the strings (color fields) that the partons stretch. This interaction modifies the whole process of fragmentation and leads to fragmentation functions that are harder than the vacuum ones. We have presented a very simplified model in which we have assumed that in the relevant energy the fragmentation process is dominated by two body decay, and we have given an explanation for large $p/\pi$ ratio and small
 quenching factor. We have also significantly
improved the value of $v_2$ from pure absorption models. 
 Further work is needed to  tune the parameters better and
 make the model more realistic, hopefully bringing it in 
even better agreement with the experimental data.

\section*{Acknowledgements} This work is partly supported by the US
DOE grant. We thank B.Kopeliovich, C.M.Ko, R.J.Fries and
T. Sj\"{o}strand for helpful comments.


\begin{thebibliography}{99}
\bibitem{comment1} The argument
in \cite{Shu_v2} was originally made for nucleae with a sharp edge
and data from the 2-body correlations. Using realistic nuclear shape with a
skin and better data, from 4-cumulants, one finds that
this statement becomes only stronger.
\end{thebibliography}
\end{document}